# Human-centric Data Dissemination in the IoP: Large-scale Modeling and Evaluation

MATTEO MORDACCHINI, MARCO CONTI, ANDREA PASSARELLA, and RAFFAELE BRUNO, CNR, Italy

Data management using Device-to-Device (D2D) communications and opportunistic networks (ONs) is one of the main focuses of human-centric pervasive Internet services. In the recently proposed "Internet of People" paradigm, accessing relevant data dynamically generated in the environment nearby is one of the key services. Moreover, personal mobile devices become proxies of their human users while exchanging data in the cyber world and, thus, largely use ONs and D2D communications for exchanging data directly. Recently, researchers have successfully demonstrated the viability of embedding human cognitive schemes in data dissemination algorithms for ONs. In this paper, we consider one such scheme based on the recognition heuristic, a human decision-making scheme used to efficiently assess the relevance of data. While initial evidence about its effectiveness is available, the evaluation of its behaviour in large-scale settings is still unsatisfactory. To overcome these limitations, we have developed a novel hybrid modeling methodology, which combines an analytical model of data dissemination within small-scale communities of mobile users, with detailed simulations of interactions between different communities. This methodology allows us to evaluate the algorithm in large-scale city- and country-wide scenarios. Results confirm the effectiveness of cognitive data dissemination schemes, even when content popularity is very heterogenous.

**ACM Reference Format:**
Matteo Mordacchini, Marco Conti, Andrea Passarella, and Raffaele Bruno. 2019. Human-centric Data Dissemination in the IoP: Large-scale Modeling and Evaluation. 1, 1 (October 2019), 25 pages. https://doi.org/10.1145/nnnnnnn.nnnnnnn

## 1 INTRODUCTION

It is commonly agreed that the vast and pervasive diffusion of devices at the edge of the network is leading to an exponential expansion of the Internet at its edges [2, 7, 43]. This expansion is proceeding at a faster pace than that of the Internet core, thanks to the diffusion of mobile personal devices, IoT devices and sensors, and other objects dispersed in the physical environment that are able to connect and communicate among themselves and with other remote entities and services.

The Internet at the edge will be mostly populated by devices that are either associated with human users or are embedded in physical objects with which humans constantly interacts. This fact will put humans at the center of the Internet system [12, 13, 29, 33, 36], requiring a paradigm shift from the traditional infrastructure-centric vision of the Internet toward a new human-centric Internet paradigm. This new paradigm has been recently named as *Internet of People* (IoP) [12, 13]. In IoP, humans and their devices will become *active actors* in the provision of new services and networking functions. Since humans (and their devices) become the center of the Internet system in IoP, the behaviour of human users should be taken into consideration in the design of future applications and services.

One the main characteristics of the scenario of IoP is that personal devices are the *proxies* that allow their users to access the vast volume of data dynamically generated by other users and "physical things" around them. In order to select and disseminate the most relevant information in this scenario, we have recently demonstrated that human cognitive schemes, known in the cognitive sciences as *cognitive heuristics* [18, 21, 22], can be directly embedded in the devices' data management and decision-making algorithms [10, 11, 30–32, 42]. Specifically, we have applied cognitive heuristics in the context of data dissemination in opportunistic networks. Opportunistic networking is a paradigm for self-organising networks, where direct contacts between devices (i.e., when devices are within the transmission range of their wireless interfaces) are exploited to exchange information and disseminate data. In particular, in this context we have used the *Recognition Heuristic* [19, 20], a simple mental scheme that helps the human brain to assess the relevance of an object in a set, based on its recognition level (an indicator of how "strongly" the object is memorised) retrieved from its memory (see Section 3 for a more detailed description of the Recognition Heuristic). In [8, 10, 42] we exploited this heuristic to quickly evaluate, at each node, the utility of storing a copy of a data item fetched from other encountered nodes during direct contacts. Simulation-based evaluations of this scheme have shown its ability to perform equally well as other state-of-the-art solutions, while requiring much less overhead in terms of exchanged messages and associated bandwidth consumption [10].

All the previous evaluations of this kind of systems have been carried out in small-scale scenarios (i.e., a few hundreds nodes and contents), while the performance of recognition-based data dissemination for opportunistic networks in large-scale realistic scenarios still needs to be assessed. Large-scale evaluations would allow us to characterise the performance and possible limitations of a data dissemination approach based entirely on local choices taken by personal devices. However, very large-scale performance evaluations pose many critical challenges. In fact, beyond a certain number of devices and/or content items, simulations, like the ones used in [10, 42], become easily unfeasible, since their complexity increases exponentially with the number of nodes and data items that are to be considered. While analytical models of the Recognition Heuristic inside a single, closed social community of users has been obtained [8], modelling reliably more complex scenarios with multiple communities, heterogeneous nodes behaviours, and heterogeneous data types, would make the analysis too complex to be tractable.

To address all these challenges, in this paper we use a *hybrid simulation* methodology. In a hybrid simulation model [28, 39, 40], the model of the original system is decomposed into separate components (submodels). Then, the interactions between the submodels are described using a conventional event-based simulation model, while the behaviour of each component is described at a higher abstraction level using a mathematical model. In the paper, the decomposition we use exploits key features of human mobility. In fact, it is well-known that humans spend most of their time moving in specific physical locations, because of the social relationships they have with other people who spend time in those locations. Examples of such location-based social communities are groups of co-workers, groups of friends, etc. Because each user is a member of multiple such communities, after some time they also move towards other groups, bridging them [6]. Therefore, we decompose an overall scenario of data dissemination across many different social communities, as follows: we separately use a model of the data dissemination occurring within each social community through a simple yet accurate mathematical model; then, we use conventional event-based simulation to model the effect on data dissemination of nodes' movements across communities. Using this decomposition, we are able to study the behaviour of the data dissemination scheme based on the recognition heuristic in very large-scale opportunistic networks. In fact, in this paper we evaluate the performance of our cognitive-based scheme up to a regional scale (in terms of geography), with hundreds of communities of mobile users, about 2.5 million of users and up to five millions of data items. This can be considered as a reasonable (even somehow far-fetched) horizon for data dissemination using exclusively direct communications between devices. Interestingly, we show that data dissemination continues to be quite efficient also at such scales. Specifically, the *hit rate* (the probability that interested nodes receive the data items they are interested in) is high also when content popularity is very heterogeneous, and there are data items that are requested by a low number of users. In particular, in the considered scenarios, it is sufficient that at least one node per community is interested in a given type of data to enable quick dissemination of data items of that type across the entire network. When content popularity drops below this threshold, data dissemination slows down in a graceful way with content popularity. Finally, we show that data dissemination is very efficient also when the sizes of local caches (used by nodes to contribute to the dissemination process) are very limited. This is an important feature to guarantee that data dissemination is not paid with excessive resource consumption.

The remainder of this paper is organised as follows. Section 2 summarises related works. Section 3 briefly presents the main concepts of cognitive heuristics and the recognition heuristic, and gives a concise description of our recognition-based data dissemination scheme. Then, Section 4 describes the hybrid simulation approach for testing the recognition-based algorithm on large-scale networks. The definition of the approach is based on the analytical model for data dissemination in single communities presented in Section 5. Section 6 presents the detailed large-scale performance evaluation of the data dissemination scheme. Finally, Section 7 concludes the paper.

## 2  RELATED WORK

The Internet of People [12, 13] is a radically new Internet paradigm that stresses the need for a change from a traditional platform- and infrastructure-centric approach to a new human-centric vision of Internet data and knowledge management. The increasing centrality and relevance played by humans (and their personal devices) [36] has been reported in several recent works that have highlighted how this fact is creating new opportunities for services at the edge of the network. Some examples of this kind are distributed sensing [41], mobile edge computing [4], mobile multimedia networks [37], and new services for smart cities [44].

In order to take advantage of the participation of the users' personal devices in the design and provision of new services in the IoP, proper communication schemes should be exploited. Opportunistic networks (ONs) are one of the forms of device-to-device (D2D) networking, which is considered an enabling technology in 5G environments in general [1], and in the IoP paradigm in particular [12, 13]. Indeed, since devices in an Opportunistic Network are usually associated with their human users, protocols in Opportunistic Networks have a close tie with the human behaviour and mobility patterns, thus presenting the characteristics required for IoP self-organising networking [12, 13]. In the context of this paper, we focus our attention to data dissemination schemes in ONs. Hereafter, we describe the main approaches presented in the literature for data dissemination in ONs.

One of the first approaches to data dissemination in ONs was the PodNet Project [26]. In PodNet data items are divided in channels of interest to which nodes are subscribed. Upon encounter, nodes exchange, according to different strategies, data items, storing in a local cache those they are not directly interested. As shown in [5], PodNet strategies suffer in mobility scenarios where nodes move across communities (like those we consider here), which are typical of human mobility.

Later on, other researchers have defined more elaborate dissemination algorithms that exploit social information about users. In ContentPlace [5], during contacts nodes exchange data items based on locally computed utility values, which reflect the individual node's estimates of the usefulness of each data item for the other members of the node's social communities. Community centrality of nodes is used in [24], and data is progressively moved on nodes with increasing centrality, to maximise the chance that those nodes will encounter other interested nodes. Social information is used also in SocialCast [14]. SocialCast also assumes that data is categorised in topics and nodes advertise their interests upon encounters. Locally computed utility values are used to decide which data to replicate on encountered nodes based on collected interests. In [46] authors propose a pub/sub system adapted to ONs. Under the assumption of a community detection algorithm, a Broker is defined for each community as the most central node. Brokers collect subscriptions in their community, and advertise interests among them. Data of interest to other communities is therefore circulated among Brokers and finally disseminated in the target communities.

Schemes not considering social information have also been proposed. Among them we mention PrefCast [27] where nodes compute a forwarding schedule to prioritise dissemination of contents that are predicted to be more requested in the near future. Other approaches for data dissemination consider solutions based on global utility functions to be solved as a global optimisation problem, where nodes' individual caches are viewed as a big, cumulative caching space, e.g. [35]. Such approaches can find an optimal solution to the dissemination process, but require global knowledge, and therefore they might be hardly applicable to ONs.

As discussed in the Introduction, this paper is part of our works on cognitive data dissemination for ONs [8, 10, 30–32]. These previous works show that simple cognitive decision-making strategies can be effectively applied in this field. In particular, in [10] we showed that mechanisms based on the recognition and take-the-best cognitive heuristics are as effective ,as other state-of-the-art solutions, while requiring considerably less overhead.

In this paper, we present a hybrid simulation study of the performance of the data dissemination approach based on the Recognition Heuristic[1].

Hybrid simulation is a technique used in the modeling and simulation field [17]. It is exploited to face problems of scalability when evaluating large systems. This makes traditional simulation studies extremely complex and expensive, in terms of both time and computational resources. On the other hand, it is in general very difficult to devise an analytical model that is able to properly describe all the relevant details of the system. Hybrid simulation exploits both simulation and analytical models trying to balance the complexity reduction, which can be achieved with analytical models, with the modeling accuracy which can be obtained with simulation. Specifically, with hybrid simulation, a complex model is analysed by decomposing it into smaller and simpler submodels. Analytical models and traditional simulation techniques (like Discrete Event Simulations) are then combined to analyse these submodels and the interactions between them, thus allowing to describe the whole system with a reduced complexity, with respect to a pure simulation study, but with a higher accuracy with respect to a pure analytical model. Hybrid simulation solutions has been applied in various fields, like computer networks [9], IoT [15], biological [25] and medical systems [3], operations research [34].

The solution described in this paper is based on the work we presented in [8]. Specifically, in [8] we developed an analytical model to analyze the behaviour of our data dissemination approach within a single community of users. Here, we extend that work by analyzing, via hybrid simulation, the performance of our dissemination approach in large scale scenarios that include hundreds of communities of users, several thousands of data topics, and millions of nodes and data items. The analytical model that we have developed in [8] is used within the hybrid simulation study to analyze each submodel in isolation.

---

[1] We refer the reader to Section 3 for a more detailed presentation of the general features of cognitive heuristics, and of the Recognition Heuristic in particular (Section 3.1).

In this paper we do not compare our approach with other solutions since: i) the Recognition Heuristic-based solution has been already successfully tested against other approaches in [10]; ii) our main goal is to develop a hybrid simulation approach that allow us to test our solution at scales (in terms of simulation area, number of nodes and data items) where other approaches become unfeasible and too complex to be studied .

## 3　THE RECOGNITION HEURISTIC AND ITS USE IN OPPORTUNISTIC NETWORKS

As in most of the related literature (see Section 2), in this study we consider a scenario where mobile nodes generate data items (these data may include data sensed by a mobile personal device from the environment) and other nodes can be interested in those items. Data items belong to high-level topics, called *channels*, to which the users subscribe. The popularity of channels (number of nodes interested in the channel) is variable across channels. The goal is to bring all the data items of a given channel to all the nodes that are interested in it. The details of the algorithm considered in the paper have been presented in [10]. Hereafter, we provide the key elements necessary to understand the analysis carried out in the rest of the paper.

In our algorithm,  each node makes use of several internal data caches. Figure 1 shows the internal architecture of these data caches on a generic node.

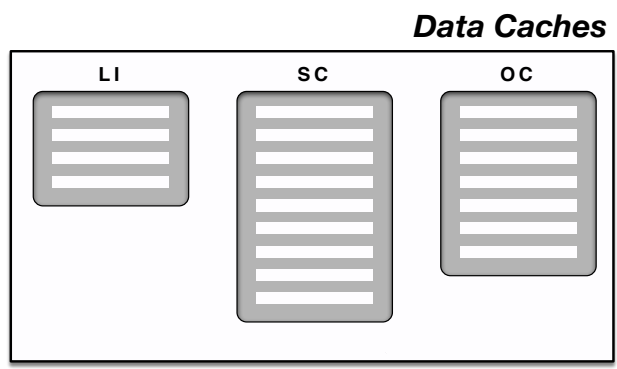

Fig. 1.  Device Data Caches

As shown in Figure 1, the data caches are:
- LI is the cache containing the *Local Items*, i.e. the items generated by the node itself.
- SC is the *Subscribed Channel* cache, i.e. the cache containing the items belonging to the channel the node is subscribed to. Data items in this cache are obtained by encounters with other peers. We assume that nodes are able to always allocate enough space for the channel they are interested in. Thus, the SC cache is assumed unlimited.
- OC is the *Opportunistic Cache*, i.e. the cache containing the objects obtained through exchanges with other nodes and belonging to channels the node is not interested in. This cache is the local storage space that a node contributes for the overall data dissemination process, beyond its particular interests. Therefore, we assume that it has a limited size. Moreover, this cache is in general much smaller than the amount of data items available in the environment.

Its content consists of the items the node believes to be the most "useful" (i.e., needed by other nodes) for a collaborative information dissemination process.

Similarly to most opportunistic data dissemination schemes, the algorithm we have proposed in [10] estimates, upon each contact, the utility for the global dissemination process that a node fetches data items from the encountered node, possibly replacing some of the data items locally stored in its OC . The original trait of our algorithm is to use the recognition heuristic to approximate the utility values of data items. The recognition heuristic is one of the cognitive decision-making processes known as *cognitive heuristics*. In Section 3.1, we briefly describe the recognition heuristic in general, while in Section 3.2 we describe how this is used in our data dissemination scheme.

### 3.1 Cognitive Heuristics and the Recognition Heuristic

Cognitive heuristics can be defined as simple rules used by the brain for facing situations where people have to solve a complex problem quickly, relying on a partial knowledge of all the problem variables, the evaluation criterion of the different possible choices is not known, and the problem itself may be ill-defined in such a way that traditional logic and probability theory are prevented to find the optimal solution. Although the general concept of heuristic is similar to what is widely used in computer science, heuristics used by human cognitive processes are formalized in a different way by cognitive sciences. We refer the interested reader to [10] for a more complete presentation of them.

In this paper, we use the *recognition heuristic*, which is one of the most popular heuristics in the cognitive science literature of the last decade [20, 23]. The recognition heuristic is based on a very simple rule. The recognition heuristic is a model of how the human brain chooses among different options based on partial knowledge about them. Specifically, instead of acquiring complete information about all options, according to the recognition heuristic, the brain chooses the option that is *recognized* with respect to those that are not. An option is recognised if it has been "seen" in the physical environment a sufficient number of times, where "seen" means that the brain has been presented with that option a sufficient number of times in the recent past. For example, in order to choose among two restaurants, one which belongs to a very well-known brand and one whose name is totally unknown, the brain would pick the former one, just because the name is well-known to it (of course, in this case, other heuristics would be applied to choose also based on quality of the food). Finally, if more options are recognised, the brain will use other heuristics (or the recognition heuristic applied to other features of the option) in order to further discriminate and identified the option to be chosen.

### 3.2 Local replacement of data items using the Recognition Heuristic

Without loss of generality, to present the algorithm we focus on the case where a tagged node encounters another node. The tagged node has to decide which data items to store in its OC, among the union of the ones it currently stores, and those available on the OC and LI caches of the encountered node.

In order to exploit the recognition heuristic to select the most relevant data items, the tagged node decides what items to store (and fetch, if they are at the encountered node) based on an estimate of their utility for the dissemination process. The utility of data items is estimated based on the following assumptions.

- the more nodes are encountered, which are subscribed to a given channel, the more relevant are the data items of that channel (and their utility increases),
- the less a data item is replicated in the network, the more useful it is to store a copy of that item.

The algorithm proposed in [8, 10] uses a two-step process to estimate the utility of a data item. In the first step, it applies the recognition heuristic to the channel of the data item, in the second it applies it to the data item itself. Specifically, according to the general recognition heuristic algorithmic model, to recognize a channel (or a data item) each node maintains a counter that counts how many times it met a device that it is interested in that channel (stores that data item). Each device devotes two caches to store this information, as shown in Figure 2.

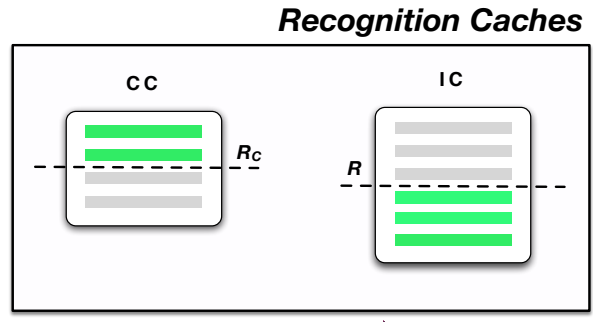

Fig. 2. Device Recognition Caches

Specifically, these caches are:
- CC is the *Channel Cache*: whenever the node meets another peer subscribed to a given channel, the channel ID is added to this cache, along with a counter. The counter is incremented every time a node subscribed to the same channel is found. Note that the storage space required to maintain this information is very limited compared with the size of data items. Thus, the size of this cache is assumed unlimited.
- IC is the *Item Cache*: similarly to the previous cache, when a new data item is seen in an encountered node, its ID is added to this cache, along with a counter. The counter is incremented every time the object is seen again. We assume that also this cache has no space limits. Even in large-scale scenarios (like the one described in Section 6), this absence of limits should not impact the search of the related IDs and counters when items are observed during an encounter, since proper retrieval techniques are available to this end. Moreover, memory requirements should remain well below the usual capabilities of standard users' devices.

Using the information stored in the caches, each node can decide that a channel (data item) is recognized if this counter reaches a given recognition threshold. In Figure 2, the channel and the item recognition thresholds are termed $R_C$ and $R$, respectively. In the experiments we have made, typical values of these thresholds varies from 2 to 25 (see Section 6). The recognition counter is decremented if the channel (data item) is not seen for a while. This is equivalent to implement a memory forgetting process in the recognition heuristic, which is known to be beneficial to improve the accuracy of the recognition itself [20]. Thus, a node considers to store a data item (either a local one, or one stored at the encountered node) if: *a*) it recognises the channel the data item belongs to, and *b*) is does *not* recognise the data item itself. All data items currently stored at the node, as well as all those stored at the encountered nodes, are analysed according to these rules. Non-recognised data items of recognised channels are candidate to be stored in the OC of the tagged node. If the candidate items are more than what can be stored, the tagged node ranks them according to their (data item) recognition level. Data items whose counter is lower (i.e., they are "farther away" from being recognised) are preferred.

## 4 HYBRID SIMULATION PERFORMANCE MODELLING

The scenario we target consists of a possibly very large number of nodes (users) that move according to realistic human mobility patterns. We also aim to simulate a possibly very large number of content items to be disseminated, which we assume to be divided into pre-defined channels. We want to track the evolution over time of the diffusion of content items, i.e. the number of nodes interested in a given channel that, at each point in time, have received the content items of that channel. This performance index, hereafter referred to as the *hit rate*, allows us to understand how well the recognition-based dissemination policy supports data dissemination as a function of the number of users, data items and channels, popularity of channels, mobility patterns.

We start by defining the nodes' mobility patterns. To this end, we consider the Home-Cell Community-based Mobility Model (HCMM), defined in [6]. This is one of the reference models in the opportunistic networking community, as it is possible to define complex mobility patterns that well reproduce known key features of human mobility, such as the distribution of contact and inter-contact patterns, social aggregation, and regularity of visits [6]. In HCMM, the space is divided into geographically separated communities. Each user belongs to just one specific community. The geographical area that contains the community of a node is called the *home cell* of the node. In this context, communities group together users that share strong social connections. Some of the nodes are allowed to travel across different communities, thus bridging them and allowing the delivery of data from one group of users to another. These nodes are called "*travellers*". Travellers have social relationships with (a subset of) the members of all communities they visit. The key parameters of the mobility model for our purposes are the number and size of communities, the number of travellers, and the probabilities of travellers moving inside their home cell or visiting an external community.

For what concerns nodes interests, we assume that each node is interested in a single channel. As it will be clear from the details of the evaluation model, the extension to the case of multiple subscriptions is straightforward, and considering single subscriptions allows us to highlight more clearly the properties of the dissemination process. For our purposes, key input parameters are: the number of channels, the number of data items per channel, the distribution of interests across nodes, the initial distribution of content items on nodes, and the size of the OC at each node.

Given this scenario, we decided to use a hybrid simulation methodology, as neither analytical models nor regular event-based simulation models suit our needs. Event-based simulation models (such as those used in [10, 31, 42]) simply do not scale beyond a certain number of nodes (a few hundreds), because the number of events increases exponentially with the number of nodes. On the other hand, analytical tractability requires that mathematical models (such as that used in [8]) abstract too many details, thus loosing in accuracy and predictive power. In [8], we have been able to obtain an accurate analytical model only for the dissemination process inside single HCMM communities, thanks to the fact that mobility of nodes inside communities is homogeneous, and the inter-contact times between nodes can be modeled as i.i.d. random variables. This model is described in details in Section 5.

To scale the model towards heterogenous scenarios with different (and possibly many) communities of mobile nodes, we leverage a hybrid simulation methodology, as follows. In the rest of this section, we make use of the symbols described in Table 1.

We use an event based simulation model only to simulate the events related to mobility of travellers, and the possible meetings between travellers while they are moving between communities. Instead, we use the analytical model presented in [8] to describe the dissemination inside a single community. Specifically, for any given content item $a$, the analytical model provides $r_t(a)$, which is the proportion of nodes in the community that hold in their shared item cache a copy of $a$ at

Table 1. List of symbols and notations

| Symbol | Definition |
|---|---|
| $a$ | Content item |
| $N_c$ | Number of nodes inside a community $c$ |
| $B$ | Size of the OC (equal for each node) |
| $r_t(a)$ | Perc. of nodes in a community holding a copy of $a$ at time $t$ |
| $r(a)$ | Steady-state value of $r_t(a)$ |
| $I$ | Set of all the items in the system |
| $I_c(t^*)$ | All the items that are available in community $c$ at time $t^*$ |
| $p(\hat{a})$ | Probability within a community for a tagged item $\hat{a}$ to be found in a community node's OC |

---

**Algorithm 1** Items selection for a traveller $n$ leaving a community $c$ at time $t^*$

1: Let $I_c(t^*) = \{a \in I | r(a) \neq 0 \text{ in } c\}$
2: Let $A$ be a multiset of cardinality $N_c \times B$
3: Fill $A$ by randomly extracting $N_c \times B$ items (potentially non distinct) from $I_c(t^*)$
4: Fill the OC of $n$ with $B$ distinct items , chosen from $A$ using a uniform random probability distribution.

---

time $t$ (after a given initial time instant), starting from any allocation of $a$ in the community (or, equivalently, the probability that any given node in the community stores item $a$). Through the model we can also identify, when a steady state exists, the constant value of $r_t(a)$ after an initial transient regime, i.e. $r(a) = \lim_{t \to +\infty} r_t(a)$. We can also identify the length of the transient regime. Therefore, when in simulation a traveller enters a community at time $t^*$, we use the current steady state value of $r(a)$ (for any data item $a$ that is possibly stored in the community nodes) as the initial condition for data dissemination in the community. We then use the model to identify a new steady state value $r(a)$, which is a function of the initial condition, and of the dissemination process of the data items brought by the traveller. The key assumption of the hybrid modeling strategy at this step is that travellers stay in a given community long enough that: i) they can "see" all data items that are interesting to them, and ii) $r_t(a)$ reaches its steady state value before they leave. Based on the results shown in Section 6.1, this is a reasonable approximation, even when the sojourn time of travellers in the communities is very short. The caches of the traveller upon exiting are populated as follows. As far as the cache containing the items of the channel the node is subscribed to, we assume that upon exiting it stores all data items of interest available in the community (i.e., for which $r(a)$ is non zero). As far as OC, the new steady state value $r(a)$ (for each possible data item $a$) is used to populate it, following Algorithm 1.

Essentially, we assume that the OC of the traveller is a random sample (with uniform probability) of all data items possibly available in the community. Specifically, we consider the set $I_c(t^*)$ of all the items (out of the set $I$ of all the items in the system) that are possibly available in the community $c$ at time $t^*$ (line 1 of the algorithm). We then produce a group of items $A$, of size $N_c \times B$, where $N_c$ is the number of nodes in the community and $B$ is the OC size (equal for each node). The set $A$ is generated as a specific realization of the population of data items in the OCs of nodes in the community, given the probability distribution obtained from $r(a)$. Thus, the probability of an item $\hat{a}$ to be in any node's OC is

$$p(\hat{a}) = \frac{r(\hat{a})}{\sum_{a \in I_c(t^*)} r(a)} \quad (1)$$

Note that, the set $A$ can contain duplicate items, as it is the juxtaposition of the realisations of the individual nodes' caches. The final set of distinct items used to fill the OC of the exiting traveller is then selected uniformly at random from $A$ (line 4).

Note that the same algorithm can also be used when multiple travellers visit the same community at the same time, as long as a $r_t(a)$ has reached a steady-state value whenever a traveller enters or exits the community. To test these aspects, we validate the accuracy of the complete hybrid simulation model in Section 6.

## 5 ANALYTICAL MODEL FOR SINGLE-COMMUNITY DATA DISSEMINATION

The goal pursued by the analysis presented in this section is to develop an analytical model describing the temporal evolution of $r_t(a)$ inside a single community of users for a generic data item $a$ belonging to a *tagged channel* $c$. In order to derive this model, we can observe that, in case of a single community, HCMM behaves like a Random Waypoint Model [6]. Therefore, nodes' mobility is homogenous, allowing to describe the nodes mobility process with iid random variables. As a consequence, we can study the replication of $a$ by considering changes in the caches of a single *tagged node*, since the same average behaviour is observed in all the other nodes of the community. Moreover, we can note that it is possible to write $r_t(a)$ as the sum of three components, i.e.

$$r_t(a) = r_0(a) + (1 - r_0(a)) \left[ P(n \text{ subs to } c) P(a \in \text{SC}) + P(n \text{ not subs to } c) P(a \in \text{OC}) \right] \quad (2)$$

In Equation 2, $r_0(a)$ is the initial replication of data item $a$ among the nodes in the community. Moreover, $P(n \text{ subs to } c)$ (i.e., the probability that the tagged node is subscribed to channel $c$) represents the popularity of $c$ among nodes in the community.

In the rest of this section, we show how to compute $P(a \in \text{SC})$ (Section 5.1) and $P(a \in \text{OC})$ (Sections 5.2, 5.3, 5.4). To this end, for each channel $c$, we make use of four distinct Markov Chains that are used to model: **i)** the presence of the generic data item $a$ in the SC of the tagged node; **ii)** the recognition level of the tagged channel $c$ in the CC of the node; **iii)** the recognition level of $a$ in the node's IC; **iv)** the presence of $a$ in the node's OC.

Although the formal definition of the state spaces of these Markov chains is reported later, for the sake of presentation clarity in Table 2 we list the main notations used in our analysis.

### 5.1 Subscribed Channel Cache

Let us consider a generic data item $a$ belonging to the channel $c$ to which the tagged node is subscribed. Then, in the Markov chain describing the status of the SC cache for that data item, state 1 is used to indicate that a copy of data item $a$ is stored in SC, while state 0 indicates that SC does not hold a copy of $a$. Let us assume that at point of time $t+1$ the tagged node encounters another node, while the previous encounter event occurred at time $t$. Intuitively, the probability $p_{0,1}$ to move from state 0 to state 1 at time $t+1$, i.e., the probability that a copy of data item $a$ is fetched by the tagged node from the node encountered at point of time $t+1$, is given by $p_{0,1} = r_t(a)$. On the other hand, $p_{0,0}$ is $1 - r_t(a)$. As the size of SC is assumed unlimited, it holds that $p_{1,0} = 0$ and $p_{1,1} = 1$. Denoting with $\psi_a^t = \{\psi_a^t[0], \psi_a^t[1]\}$ the state vector of the Markov chain at time step $t$, and $P = \{p_{i,j}\}^2$, $i, j = 0, 1$, its transition matrix evaluated at time $t+1$, the evolution of the SC Markov Chain is as follows: $\psi_a^{t+1} = \psi_a^t P$.

---

[2] Note that the $p_{i,j}$ values depend on the parameter $t$, but for the sake of notation simplicity we omit it.

Table 2. List of mathematical notations

| Symbol | Definition |
|---|---|
| $N$ | Number of nodes of the network |
| $M$ | Number of items in the network |
| $C$ | Number of channels in the network |
| $a.c$ | Channel to which data item $a$ belongs |
| $r_t(a)$ | Fraction of nodes storing a copy of $a$ |
| $Pop(c)$ | Probability for a node to be subscribed to channel $c$ |
| $\psi_a^t[1](\psi_a^t[0])$ | Prob. that data item $a$ is (is not) in SC at time $t$ |
| $\nu_c^t[i]$ | Prob. that channel $c$ has a recognition level $i$ in CC at time $t$ |
| $\upsilon_a^t[i]$ | Prob. that data item $a$ has a recognition level $i$ in IC at time $t$ |
| $\varphi_a^t[i]$ | Prob. that data item $a$ is stored in OC at time $t$ with a rec. level $i$ |
| $R_c$ | Threshold for channel recognition |
| $R$ | Threshold for data item recognition |
| $B$ | Max. number of slots available in a OC |

## 5.2 Channel Recognition

As described in Sec. 3, CC stores the number of times the tagged node has encountered another node subscribed to the tagged channel. Thus, the state $i$ of its Markov Chain is the number of times this has happened, which is the recognition level of the channel. The transition matrix is defined as follows

$$p_{i,j} = \begin{cases} Pop(c) & \text{if } j=i+1, i \in [0, R_c-1] \\ \alpha^i(1-Pop(c)) & \text{if } j=i-1, i \in [1, R_c] \\ 1-((1-\alpha^i)Pop(c)+\alpha^i) & \text{if } j=i, i \in [1, R_c-1] \\ 1-Pop(c) & \text{if } i,j=0 \\ 1-\alpha^i(1-Pop(c)) & \text{if } i,j=R_c \\ 0 & \text{otherwise} \end{cases}, \quad (3)$$

The probability that $i$ increases at an encounter is the probability that the encountered node is interested in the channel, i.e. the channel popularity $Pop(c)$. We drop the recognition level by 1 when the encountered node is not interested in the channel (which happens with probability $1 - Pop(c)$). This is weighted by a factor $\alpha^i$, with $0 \leq \alpha \leq 1$. Variable $\alpha$ replicates the phenomenon of *forgetting*. As reported in the cognitive science literature [16, 38, 45], the memory trace associated with an item decays over time exponentially. However, repeated expositions to the item reinforce its memory trace, thus slowing down the forgetting process. For these reasons, factor $\alpha^i$ decreases as the recognition level increases. Therefore, we drop recognition of channels that are not popular (anymore), but we do so more conservatively for very popular channels, since they have been observed more times, achieving an increased memory strength. The probability of remaining at the same recognition level is the complement of the first two transitions. Special cases are state $i=0$, where the recognition level cannot decrease, and state $i=R_c$, where the recognition level cannot increase.

The state vector of the CC chain is denoted by $v_c^t$, and the initial state at the beginning of the process evolution is $v_c^0 = \{1, 0, \ldots, 0\}$.

### 5.3 Item Recognition

Conceptually, the recognition process of data items is similar to the channel recognition. In fact, IC maintains a list of the observed data items and their recognition levels, defined as the number of times they have been observed in encountered nodes. Thus, the state $i$ of the Markov chain used to describe the status of the IC for the generic data item $a$ belonging to channel $c$ represents its recognition level, while state 0 represents the condition of data item $a$ not having been observed, yet. Considering that the probability that a tagged node finds a copy of data item $a$ in the caches of another node encountered at point of time $t+1$ depends only on its replication level $r_t(a)$, the transition probability matrix of this Markov chain is as follows

$$p_{i,j} = \begin{cases} r_t(a) & \text{if } j = i+1, i \in [0, R-1] \\ \gamma^i(1-r_t(a)) & \text{if } j = i-1, i \in [1, R] \\ 1-[r_t(a)(1-\gamma^i) + \gamma^i] & \text{if } j = i, i \in [1, R-1] \\ 1-r_t(a) & \text{if } i, j = 0 \\ 1-(1-r_t(a))\gamma^i & \text{if } i, j = R \\ 0 & \text{otherwise} \end{cases}, \qquad (4)$$

where $\gamma$ ($0 \leq \gamma \leq 1$) is the discount factor for the item recognition level. In the following, $v_a^t$ is the state of the IC chain, with initial state $v_a^0 = \{1, 0, \ldots, 0\}$.

### 5.4 Opportunistic Cache

Differently from the other caches, OC has a limited size $B$. As described in Sec. 3.2, only copies of data items belonging to recognized channels can be stored in the OC cache. However, given that OC can contain only a subset of available data items, a replacement policy is needed. We recall that our modified *Take-the-Best* replacement policy gives a higher priority to data items with lower recognition levels. In case of ties, the data items found in the caches of the encountered node are preferred to the ones already in the OC cache of the tagged node. To tackle the problem of modeling the evolution of OC, we represent this cache as a *queuing network*, as shown in Figure 3. In this queuing network, a sub-queue $i$ stores all data items that are in the OC and have the same recognition level $i$. The sub-queue 0 is a virtual queue that contains all the data items that are outside of the OC, independently of their recognition levels. Then, the $i$th element ($i > 0$) in the state vector $\varphi_a^t = \{\varphi_a^t[0], \varphi_a^t[1], \ldots, \varphi_a^t[R]\}$ represents the probability that the generic data item $a$ is in the OC cache with recognition level $i$ at time $t$. It is also important to note that each individual sub-queue has not a fixed size, but we must ensure that the sum of the numbers of data items stored in these sub-queues (excluding sub-queue 0) is lower or equal to $B$.

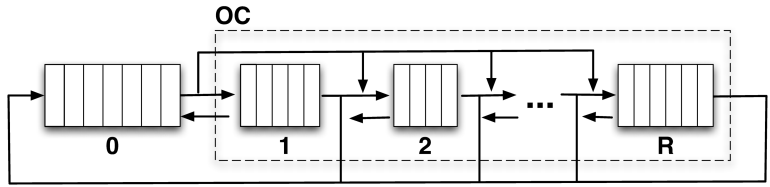

Fig. 3. Queuing network modeling the OC cache.

To analyse the replacement policy for OC, we split the problem into two simpler sub-problems. First of all, we model the reordering of the data items stored in the OC cache due to changes in their recognition levels after an encounter event. For instance a data item that was initially stored in sub-queue $i$ should be moved to sub-queue $i+1$ if its recognition level is increased upon an encounter event. Since this process involves only an internal reordering of stored data items, no data items are dropped. The second step in the analysis takes into account that new data items fetched by the caches of the encountered node may enter the OC of the tagged node at a sub-queue that depends on their recognition level. Due to cache size constraints, some of these data items could not be allowed to enter the OC cache, or some data items already stored in OC could be removed to let new data items enter the OC cache. In addition, in both steps we consider the possibility that data items in the OC are moved to sub-queue 0 – i.e. are dropped – if their channel is not recognized anymore after the encounter occurring at point of time $t+1$. In the following we separately describe these two modeling phases.

**Step 1.** Let us introduce an auxiliary Markov chain, whose state vector $\varphi'_a = \{\varphi'_a[0], \varphi'_a[1], \ldots, \varphi'_a[R]\}$ represents the probability that the generic data item $a$ is in the OC cache with recognition level $i$ *after* the encounter event, but *before* new data items are inserted in the OC. Then, we have that $\varphi'_a = \varphi^t_a P'$, where $P'$ is the transition probability matrix of the auxiliary Markov chain modeling the data item reordering. Since no data items enter OC in this phase, we have that $p'_{0,0} = 1$. However, a data item could be removed from OC if the channel it belongs is not recognised anymore, which happens with probability $(1 - v^{t+1}_a[R_c])$, or the recognition level of the data items becomes 0. In other words

$$p'_{i,0} = 1 - v^{t+1}_a[R_c], i \in [2, R] \tag{5}$$

$$p'_{1,0} = (1 - v^{t+1}_a[R_c]) + v^{t+1}_a[R_c]\gamma(1 - r_t(a)) \tag{6}$$

Formula (6) can be explained by noting that a data item of a recognized channel can change its recognition level from one to zero only if it is not in the caches of the encountered node, and the tagged node applies the discount factor $\gamma$ to the data-item recognition level stored in IC (see formula (4)). Following a similar line of reasoning as in (6) we can compute the probability that a data item already in the OC moves to either backwards or forwards sub-queues. More formally we have that

$$p'_{i,i-1} = v^{t+1}_a[R_c]\gamma^i(1 - r_t(a)), i \in [2, R] \tag{7}$$

$$p'_{i,i+1} = v^{t+1}_a[R_c]r_t(a), i \in [1, R-1] \tag{8}$$

On the other hand, the probability of remaining in the same sub-queue after the encounter event at point of time $t+1$ is simply given by the complement of the sum of the probabilities of moving backwards or forwards and to leave the OC cache. More formally, it holds that

$$p'_{1,1} = 1 - (p_{1,0} + p_{1,2}) \tag{9}$$

$$p'_{i,i} = 1 - (p_{i,0} + p_{i,i-1} + p_{i,i+1}), i \in [2, R-1] \tag{10}$$

$$p'_{R,R} = 1 - (p_{R,0} + p_{R,R-1}) \tag{11}$$

We now exploit the knowledge of the state vector $\varphi'_a$ to compute the new average number of data items in each sub-queue $i$, say $B'_i$, after the internal reordering, which is simply given by

$$B'_i = \sum_{a=1}^{M} \varphi'_a[i] . \tag{12}$$

**Step 2**: To derive the final status of the OC cache, i.e. $\varphi_a^{t+1}[i]$, in this step we first compute the average number $N_{0,i}$ of new data items that are eligible to enter OC at sub-queue $i$. Then, we compute the number $F_i$ of available free slots at each sub-queue $i$ of the OC. The $F_i$ value is the key parameter we need to compute the probability that a new data item is either discarded or cached, and the probability that an old data item stored in OC is removed. $N_{0,i}$ is computed as follows:

$$N_{0,i} = \sum_{a=1}^{M} (1 - r_0(a))\, r_t(a) v_a^{t+1}[R_c] v_a^t[i-1] \varphi_a^t[0] \,. \tag{13}$$

In this definition, we can observe that a new data item not already stored in the OC of the tagged node is eligible for entering OC in the sub queue $i$ **iff** (i) it is not one of the items generated by the tagged node itself (that happens with probability $1 - r_0(a)$), (ii) it is outside the OC at time $t$ (probability $\varphi_a^t[0]$), (iii) it is stored in the caches of the encountered node (probability $r_t(a)$), (iv) the channel it belongs to is recognized (probability $v_a^{t+1}[R_c]$), (v) and its recognition level before the encounter event was $i-1$ (probability $v_a^t[i-1]$).

It is important to remind that $N_{0,i}$ expresses the number of new data items that can be potentially copied in the OC cache. However, the actual number of new data items that are copied in OC will depend on the number of free slots. More precisely, let us denote with $F_i$ the *maximum* number of free slots that new data items can occupy in the sub-queue $i$ of the OC cache. It holds that

$$F_i = B - \sum_{j=1}^{i-1} B_j^{t+1} \,, \tag{14}$$

with $F_1 = B$. Indeed, data items (both new or old) with recognition level equal to one have the highest precedence and they can use the entire OC. On the contrary, data items with recognition level equal to $i$ ($i > 1$) can use only the part of OC not used by data items with lower recognition levels. It is also important to point out that in formula (14) we must use the $B_i^{t+1}$ value because it provides the size of sub-queue $i$ in OC *after* completing the internal reordering, the insertion of new items and the removal of old items. However, it is quite straightforward to observe that $B_i^{t+1}$ is simply given by

$$B_i^{t+1} = \min(N_{0,i} + B_i', F_i) \,. \tag{15}$$

Formula (15) can be explained by noting that if $N_{0,i} + B_i' > F_i$, then there are enough free slots in OC for all the new items that could enter at level $i$ and for the old items that are already at level $i$ after the reordering. On the other case, some new items will be discarded and/or some old items will be dropped till only $F_i$ slots are occupied, as discussed later. Now, by using formulas (14), (15) and the initial condition $F_1 = B$ we can iteratively compute all remaining $B_i^{t+1}$ and $F_i$ values.

Finally, to compute the state vector $\varphi_a^{t+1}$ we introduce an auxiliary transition probability matrix $P$ such that $\varphi_a^{t+1} = \varphi_a' P$. As observed in formula (13), a new data item is eligible for entering OC at recognition level $i$ with probability $r_t(a) v_a^{t+1}[R_c] v_a^t[i-1]$. Remember that, when we need to pick only a subset of the $N_{0,i} + B_i'$ data items that could possibly stay in sub-queue $i$, we prefer new items over items that were previously in the OC. Therefore, that new data item will certainly be fetched by the tagged node if $N_{0,i} \leq F_i$, otherwise only a fraction $F_i/N_{0,i}$ will be fetched. Thus, the probability $p_{0,i}$ that a new item effectively enters the OC cache can be expressed as

$$p_{0,i} = \begin{cases} r_t(a) v_a^{t+1}[R_c] v_a^t[i-1] & \text{if } N_{0,i} \leq F_i \\ r_t(a) v_a^{t+1}[R_c] v_a^t[i-1] \frac{F_i}{N_{0,i}} & \text{otherwise} \end{cases} \tag{16}$$

On the other hand, an old data item that was already stored in the OC cache should be removed if the new data items have consumed all (or most of the) free slots. More precisely, we have that

$$p_{i,0} = \begin{cases} 0 & \text{if } N_{0,i} + B'_i \leq F_i \\ 1 & \text{if } N_{0,i} > F_i \\ 1 - \frac{F_i - N_{0,i}}{B'_i} & \text{otherwise .} \end{cases} \quad (17)$$

Formula (17) indicates that in case $N_{0,i} + B'_i \leq F_i$ there is no need of replacing data items because the free slots can accommodate both new and old data items. On the other hand if $N_{0,i} \leq F_i$ a fraction $1 - \frac{F_i - N_{0,i}}{B'_i}$ of old data items has to be removed from the OC cache. The rest of the transitions of matrix $P$ are clearly as follows:

$$p_{i,j} = \begin{cases} 1 - \sum_{k>0} p_{0,k} & \text{if } i = j = 0 \\ 1 - p_{i,0} & \text{if } j = i, i \neq 0 \\ 0 & \text{otherwise .} \end{cases} \quad (18)$$

## 5.5 Steady-state Analysis Inside a Single Community

In [8] we have presented a comprehensive analysis of the transient state of data replication based on the model. In this section, we only point out that the model is able to accurately follow the simulation results. Specifically, for the purpose of this paper it is of particular interest to focus on the steady state properties, as shown in Figure 4. The same findings reported in this figure are valid for other configurations and we refer the reader to [8] for the complete set of the model validation results. Results are obtained with a network composed of 45 nodes, 3 different channels with 99 data items each (297 data items in total). Channels' popularities are skewed and they follow a Zipf distribution with parameter 1. Nodes move according to a Random Waypoint model in a square area of side 1 km. Figure 4 reports the replication levels over time in the nodes' OCs of each tagged item of the 3 channels over time. The model is compared with the results of a conventional simulation. In this figure, the results of the most popular channel are presented in the leftmost plot, the mid popular channel is presented in the central plot, while the least popular channel's results are in the rightmost plot. From these results, we can specifically highlight two features of the system. The first one is that the replication level in the steady state is the same for all channels. As shown by the figures, the model is able to correctly account for this feature. Furthermore, the model also allows us to accurately predict the evolution of the dissemination process during the transient phase, and the time when the steady-state value of the replication function is achieved.

## 6 HYBRID SIMULATION VALIDATION AND LARGE SCALE EVALUATION

In this section we evaluate the performance of the recognition-based data dissemination policy in large-scale scenarios.

Based on the results shown in Section 5.5, we instantiate Algorithm 1, by setting an identical steady-state probability $r(a)$ for each item $a$ available in the community when an event occurs (remember that events are traveller entering or leaving a community). We first validate the hybrid simulation model in small scale against legacy simulation results. Then, we use hybrid simulation to study the system in larger settings, with wider areas and greater number of users. These settings have the geographical extension and the population size of a urban- and a regional-scale scenario, respectively. Note that in the following we mainly study the transient phase of the *entire* system (i.e., not of individual communities), i.e. the phase before the convergence of the hit rate. This is clearly different from the transient phase of dissemination in each specific community, which was highlighted in [8] .

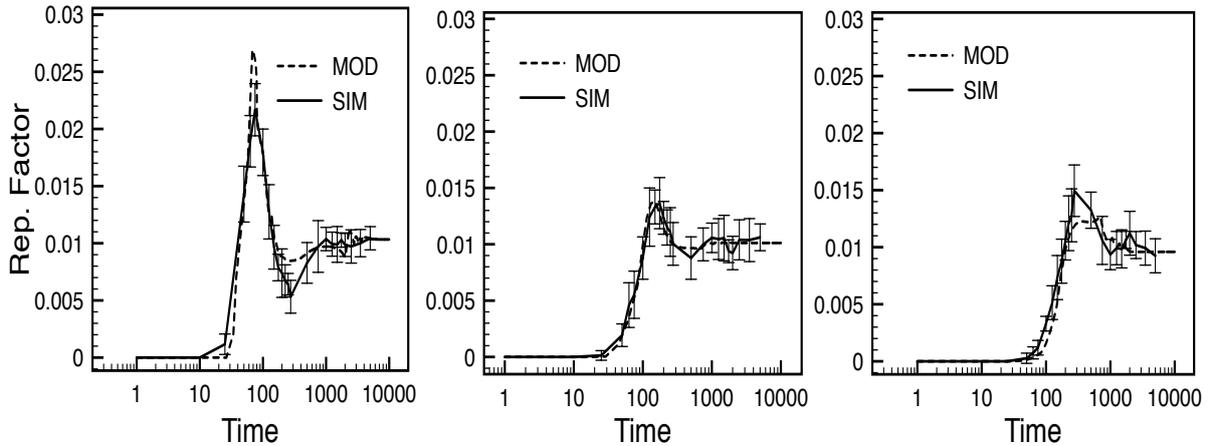

Fig. 4. $r_t(a)$ for $R_c = 5$, $R = 5$ and $B = 3$.

## 6.1 Hybrid Model Validation

In the following, all data items are generated at the beginning of the simulation, and are initially distributed uniformly at random among nodes in the network. Each node subscribes to one channel only at the beginning of the experiment. Nodes subscriptions inside each group are skewed, and are distributed according to a Zipf distribution with parameter 1. Moreover, interests are rotated, so that the most popular channel in a group is the second in another and the third one in the other, and so on. The results shown in the following figures are obtained using the simulation parameters reported in Table 3.

Table 3. Validation - Main simulation parameters

| Simulation Parameters | |
|---|---|
| Simul. Area | 1000 m × 1000 m |
| Transm. range | 20 m |
| Numb. of Communities | 3 |
| Numb. of Nodes | 45 (15 per comm.) |
| Numb. of travellers | 6 (2 per group) |
| Numb. of channels | 3 |
| Items per channel | 99 |
| Simulation time | 125000 s |

Results from both conventional and hybrid simulations are the average over 10 replicas. Unless otherwise stated, we do not show confidence intervals, since they are very small, and showing them would compromise the readability of the figures. In each replica we use an independent synthetic mobility trace obtained from HCMM with the same configuration. The hybrid simulation is designed by assuming that the spreading of data within a community reaches a steady state before a traveller exits the group. This condition is met whenever the mean sojourn time of the traveller in a community is long enough. Therefore, in order to analyse the effects of sojourn times on the assumptions of the hybrid simulation, we tested the system using different mean sojourn times. Results are reported in Figure 5. For both the hybrid and conventional simulation, we used a OC size $B = 5$, while the results of the conventional simulation reported in the figure are obtained with a Channel Recognition Threshold $R_c = 10$, and Item Recognition Threshold $R = 10$. We analyse in Figure 5 the results varying these parameters. Clearly, the two latter parameters are not relevant for the hybrid simulation, since this model assumes that the recognition of channels and items reaches a steady state before any event happens. In the figure, hybrid simulation results start at a time different from 0, since this is the time step when the first event (i.e., the exit of a traveller from a community) is registered in the simulation.

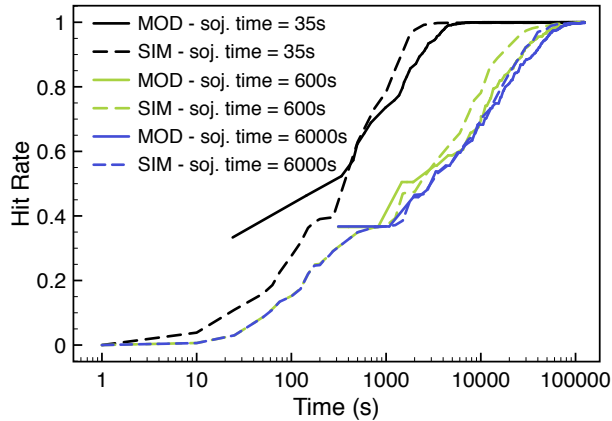

Fig. 5. Comparison of hybrid simulation and complete simulation results, with different mean sojourn times.

Looking at Figure 5, it is possible to note that there are only minor differences between the hit rate obtained by the hybrid simulation and the ones resulting from the legacy simulation, when travellers have sojourn times of 6000 and 600 seconds, respectively. This is a remarkable result, since the two sojourn times differ by one order of magnitude. When using a very short sojourn time (just 35 sec.), differences are more marked. Nonetheless, they still remain limited, with the two hit rates increasing with close and similar trends in both types of simulation. A traveller's mean

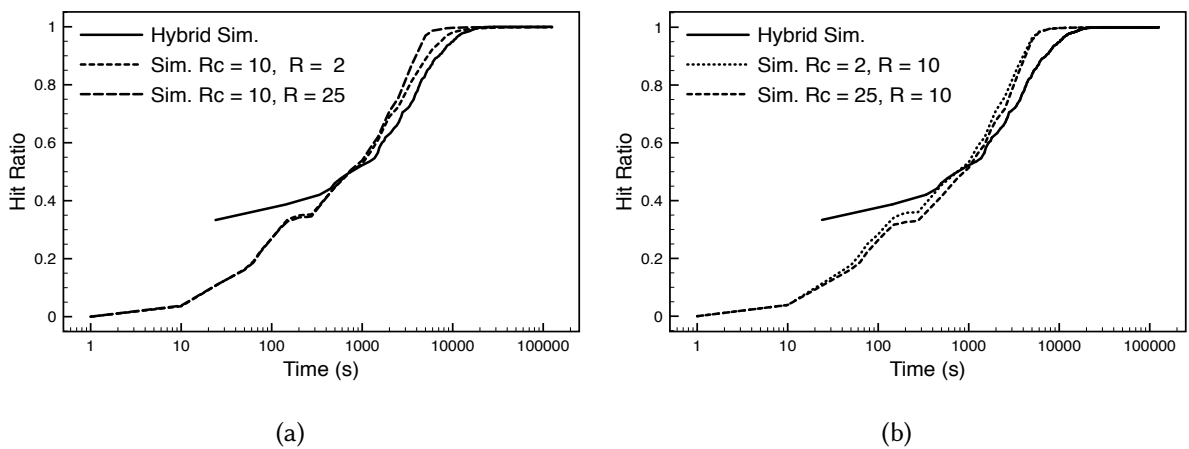

Fig. 6. Comparison of hybrid simulation and complete simulation results, with different item (a) and channel (b) recognition thresholds. Mean sojourn time = 35 s

sojourn time of 35 seconds is extremely short, and this is an interesting configuration as it stresses the assumptions of the hybrid simulation model. Therefore, in Figure 6 we further investigate the behaviour of the hybrid simulation with a mean sojourn time of 35 seconds. Specifically, we use values that represent extreme cases for the Channel and Item Recognition Thresholds in the legacy simulation. Figure 6a shows the results obtained by keeping the $R_c$ fixed to the value of 10 and by varying the value of $R$. Note that the hybrid simulation starts with a higher Hit Ratio. This happens because the hybrid simulation assumes that the nodes in the community are able to retrieve all the items of their subscribed channels before the departure of a traveller. Afterwards, the hit rate in the conventional simulation model become higher. This is more evident for larger values of $R$, because data items take longer to be recognised. Note however, that the hybrid simulation model well follows the trend of increase of the hit rate.

Table 4. Urban and Regional scale simulation parameters

|  | *Urban Scale* | *Regional Scale* |
|---|---|---|
| Simul. Area | 1,000 m × 1,000 m | 10,000 km$^2$ |
| Transm. range | 20 m | 20 m |
| Numb. of Communities | 50 | 250 |
| Numb. of Nodes | 10,000 (200 per comm.) | 2,500,000 (10,000 per comm.) |
| Numb. of travellers | 2,450 (49 per comm.) | 6,250 (25 per comm.) |
| Numb. of channels | 50 | 10,000 |
| Items per channel | 100 (def.), 200 and 400 | 500 (5,000,000 in tot.) |
| Simulation time | 125,000 s | 125,000 s |

In Figure 6b the value of $R$ is fixed to 10 and two different values of $R_c$ are used. The general behaviour is similar to that of the previous experiment, with differences between the hybrid and conventional simulations due to different recognition levels. It is interesting to note that the highest value of $R_c$ modestly affects the Hit Ratio, by slightly slowing it down, as channels need more time to become recognized and have their items diffused in the system.

## 6.2 Urban-Scale Evaluation

In this section, we present a series of experiments made with the hybrid simulation approach in a much larger scale than that of the previous experiments, in order to test the recognition-based dissemination scheme in a urban area scale. The experimental parameters used in this section are reported in Table 4. Events about the travellers' movements are obtained with the HCMM model. We consider that each group has one outgoing traveller toward each of the other communities.

Specifically, the following experiments analyse a scenario like the one of a urban area, with a considerably large number of nodes, in the order of thousands of devices. In fact, we scale up to 10,000 nodes. The number of channels and communities are one order of magnitude greater than that used in the validation study (see Section 6.1). We still assume that subscriptions within each community follow a Zipf distribution with parameter 1. Moreover, subscriptions are rotated between groups, also in this case.

Figure 7 shows the trend of the Hit Ratio over time for the most, mid and least popular channels within a tagged community. We focus on a specific community instead of the entire set of nodes because overall all channels have the same popularity, while in each community they are not equally popular. In this experiment, the size of the OC is sufficient to hold 10 data items, and there are 100 items per channel. Despite the fact that the most popular channel has 50 times more subscriptions than the least popular one, the difference in hit rate is very limited. In particular, there are no differences between the curves of the most and mid popular channels, while the least popular channel seems to converge more slowly that the previous ones. However, note that 95% confidence intervals for the most (depicted in red) and least (blue) popular channels always overlap, sometimes including the other curve. This is expected, as the steady state value for distribution of data items on nodes ($r(a)$) is equal among channels. This fact highlights one of the positive features of the recognition policy, that gives priority in the OC cache to less recognised, and thus less diffused, data items, obtaining a final fair diffusion of all the items, regardless of their channels' popularities.

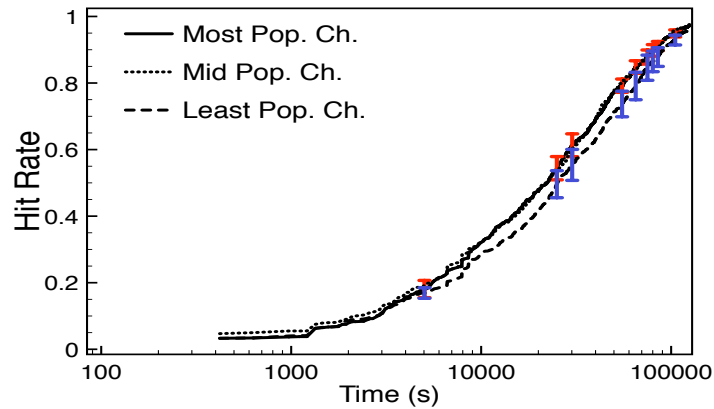

Fig. 7. Hybrid simulation results for the most, mid, and least popular channels within a tagged group

In the experiments reported in Figure 8a, we analyse the impact of the OC size on the hit rate considering the entire network (not a single community). These results are obtained assuming that each channel has 100 items and that only 2 of them are initially available within each community. This is to test the capacity of our policy to spread data items across different communities. As one could expect, the greater the size of the OC, the faster the convergence of the hit rate to its maximum value. While with an OC of 25 slots (i.e. only 0.5% of the total number of data items) a 100% hit rate is reached after nearly 60,000 sec. of simulation, with an OC of 10 slots the system stabilises at the end of the simulation. When the OC is very small, i.e. 3 slots, the hit rate remains below 60% even after 100,000 seconds of simulation. Finally, Figure 8b presents the different behaviour of the system when the number of items per channel is increased. Specifically, the system is tested with a total number of items in the system equal to 5000 (100 items per channel), 10000 (200 items per channel) and 20000 (400 items per channel). The size of the OCs is fixed to 10. The converge of the Hit Ratio is reduced when a larger number of items is involved. Larger sets of items, in fact, require more time to be spread between all the communities, thus slowing down the trend of the Hit Ratio. However, it is worth noting that the difference among the curves is proportional to the total amount of items available in the system.

### 6.3 Regional Scale Evaluation

In this section, we extend the analysis of our system to a much larger scale scenario. In this case, the simulation area and population are comparable with that of the county of San Diego, CA, the Gifu Prefecture in Japan, or with that of a small nation like Jamaica. In this context, nodes are grouped in 250 large communities of 10,000 devices each, reaching a total population of 2,500,000 users. In order to have more realistic settings, we consider a very large number of channels (10,000), with a large amount of total available items, i.e. 5,000,000 items (500 per channel). Subscriptions are skewed at the level of the *entire population*, following a Zipf distribution of parameter 1. As a consequence, there is a very long tail, with numerous channels having only a few tens of nodes subscribed to them, while the most popular channels can count on tens of thousands of subscribers. Nodes are assigned to communities according to a uniform distribution. In such a scenario, we cannot expect to have each community connected to all the others, like in the previous experiments. Specifically, each group has only a small number of outgoing travellers (25, in the experiments), connecting it to other communities. As in the experiments of the previous sections, each traveller commutes toward another group only. In this case, the destination community of a traveller is chosen at the beginning of the simulation on a geographical basis. Specifically, travellers have a skewed probability (modeled as a Zipf distribution of parameter 1) of choosing another group as their destination: the farthest a group, the lowest the probability to be connected with it. The main parameters used in the simulations are summarised in Table 4.

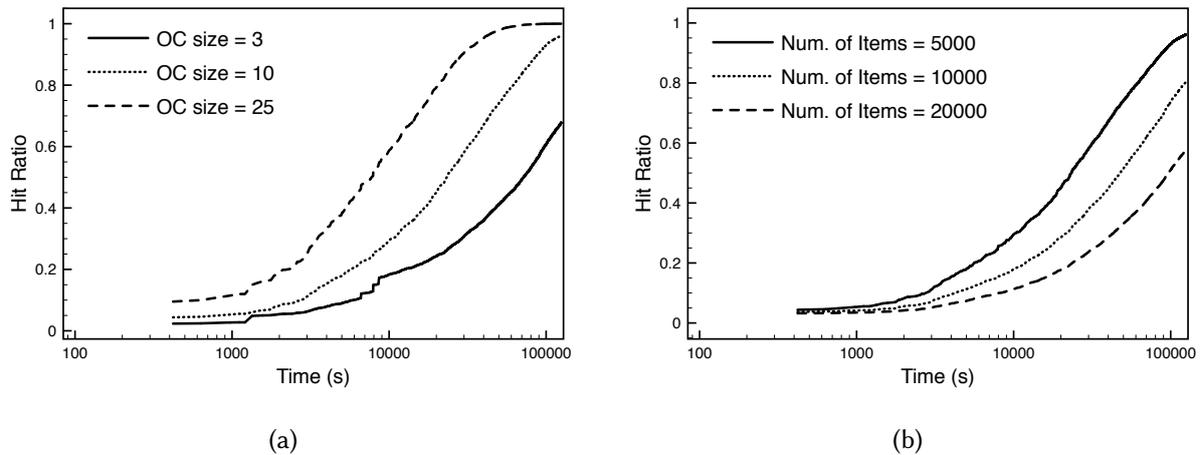

Fig. 8. Hybrid simulation results, with different sizes of the OC (a), and with different total amounts of items available in the network (b).

Due to the high skewness in the channels popularities, in this experiment we do not expect that all the channels are able to achieve the same hit rate at the end of the simulation. Our goal is to investigate the behaviour of the system and its ability to deliver data with such a highly heterogeneous data popularity, under strict limits in terms of OC size (compared to the number of channels and items), number of travellers (with respect to the number of communities and users) and conditions that make the dissemination difficult, like a wide geographic area, and a very large number of users to be served. In the experiments, we evaluate the system using two different scenarios for the initial data generation, which affect the overall dissemination process. In the first scenario (hereafter termed as Scenario 1), the data of a channel is generated by the nodes that are also subscribers of that channel. In the second scenario (Scenario 2), the data of each channel is initially placed uniformly at random across all the nodes in the system. Figures 9a and 9b report the results for these two scenarios, obtained by the system with an OC size of 5,000 slots (i.e., the 0.1% of the total number of items). Generally, the dissemination in the first scenario proceeds more rapidly than in the second one, since, in the latter case, data is more dispersed, with respect to users interests, and then it requires more time to reach the interested users. In both scenarios, all channels up to the 1000-th most popular one achieve a 100% hit rate at the end of the simulation (90% for the 1000-th most popular one in the second case) Note that the most popular channel has more than 250,000 subscribers (ca. 10% of the users), compared to just a little more than 250 subscribers of the 1000-th channel (almost 0.01% of the users). Therefore, the system is able to fairly serve the users who are interested in channels with extremely different levels of subscriptions on a regional scale. We can also note that there is a very high difference between the 1000-th subscribed channel and the 2500-th channel, much greater than that existing between the 1000-th channel and the 1st one, suggesting that a phase transition occurs in the behaviour of the system. This is due to the fact that the 1000-th least popular channel is nearly the last channel with at least one subscriber per community (on average). In fact, some channels have a number of subscribers that is lower than the number of communities. Whenever nodes in a community cannot observe any other device subscribed to a given channel (and this is the case for the less popular channels), this channel is prevented to become recognised in that community. As a result, its data items cannot be replicated inside the OCs of the nodes of that community. Although this has no effect on the hit rate of nodes of that community, this limits the overall dissemination of those data items, and impacts other communities where subscribers are present. The specific hit rate achieved by subscribers of those channels clearly depends on how travellers connect the different communities.

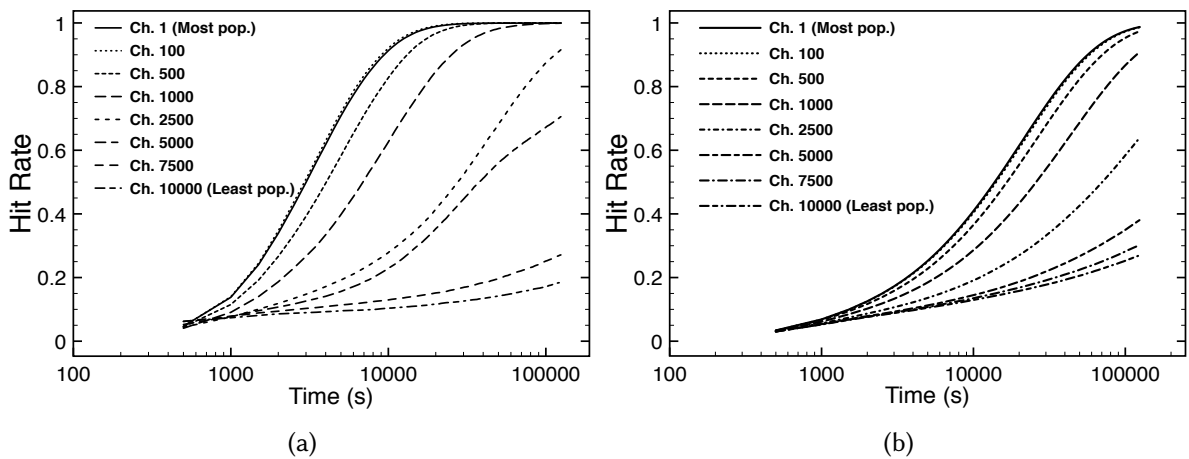

Fig. 9. Results for channels with various popularities, OC size = 5000. (a) Scenario 1 (b) Scenario 2.

These effects are particularly evident for the least popular channels. These channels are characterised by a very low number of subscribers (less than 30 users), highly dispersed among the available communities (just one subscriber every ten communities for the very least popular channels). These are very extreme conditions. In order to have a deeper understanding of the behaviour of the system with respect to less dispersed distributions of subscriptions, in the next set of experiments we analyse the cognitive-based solution when only channels with a minimum number of users per community are present in the network. Specifically, we use a setting where we keep only the users and items of channels having subscribers on at least the 25% of the communities. Note that the least popular channels of this setting are already in the long tail of the channels popularity distribution, each having 0.0025% of the total number of nodes as subscribers. Results are shown in Figures 10 and 11. Specifically, Figures 10b and 11b are obtained with the new setting of channel distributions, while Figures 10a and 11a are those referring to the system with all the channels. In both scenarios, the hit rate for all the channels is increased in the new setting. This effect is particularly significant for the least popular channels in Scenario 1 and for all the channels in Scenario 2. In both cases, data is dispersed among communities, and requires more effort to be delivered. Thus, these channels receive the most noticeable advantage from the absence of extremely poorly subscribed channels.

The previous results were obtained using the channel recognition mechanism inside the communities. In the final experiment we test whether removing the channel recognition inside communities could be beneficial for the system. The comparison is presented in Figure 12, where results are the average hit rate among all the channels, and are obtained with the same configuration used in Figures 10 and 11. In both Scenario 1 and 2, the solutions without the channel recognition increase the hit rate at a lower pace than the versions using it. In fact, without the channel recognition, data items pass through communities where they are not requested. As a result, these items could compete for entering the caches of travellers, even when they are not useful for the overall dissemination process. The usage of the channel recognition mechanism allows to have a more focused dissemination strategy, resulting in a faster diffusion of data items toward their interested users.

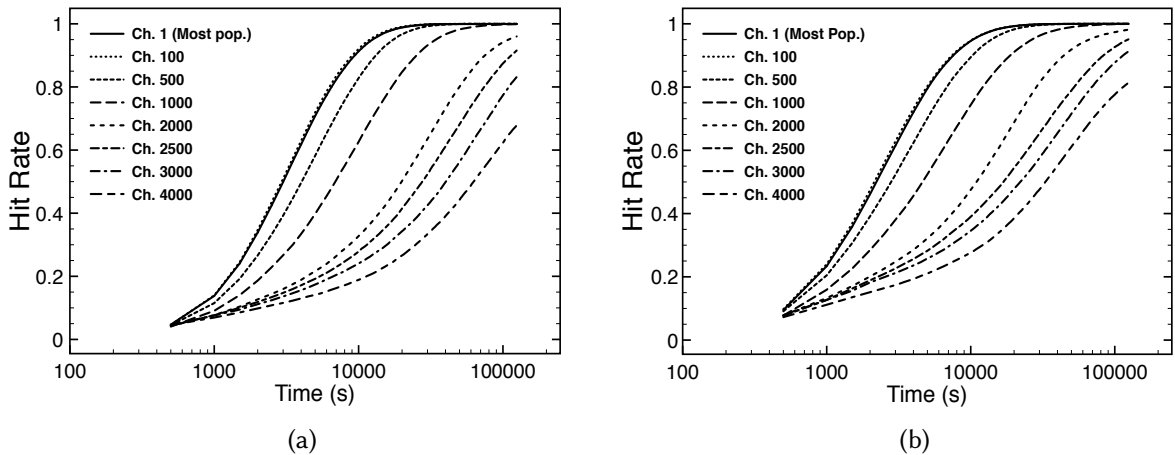

Fig. 10. Scenario 1. (a) Results with all the channel (b) Results with channels having subscribers on at least 25% of the communities

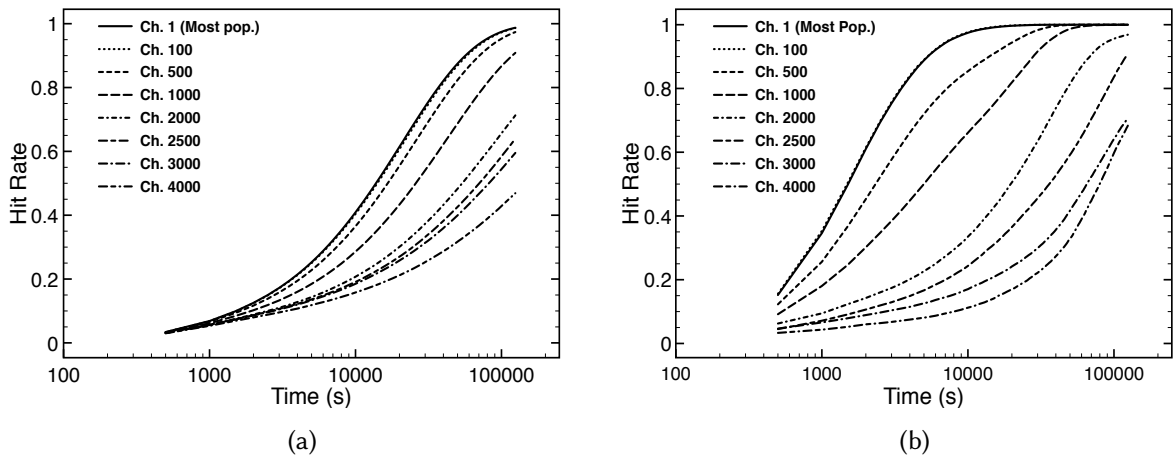

Fig. 11. Scenario 2. (a) Results with all the channel (b) Results with channels having subscribers on at least 25% of the communities

## 7 CONCLUSIONS

In this paper we have considered a data dissemination scheme for opportunistic networks where nodes embed human cognitive processes to decide how to replicate data items upon contacts. This approach is based on the general ideas that, in future IoP systems (i) self-organising networks based on devices at the edge will be a key element, and (ii) data-centric services running on personal mobile devices need to incorporate algorithms that closely match the behaviour of their human users in the physical world. The main goal of the paper was to assess the behaviour of the cognitive-based data dissemination scheme in large-scale configurations. To this end, neither purely simulation-based evaluation methods nor analytical models are sufficient. Therefore, we have applied a hybrid simulation methodology, using an analytical model to characterize the system behaviour at the community level to reduce the complexity of the simulation model. This allows us to scale up the evaluation scenarios by orders of magnitude with respect to what we could obtain with conventional simulation, without losing accuracy.

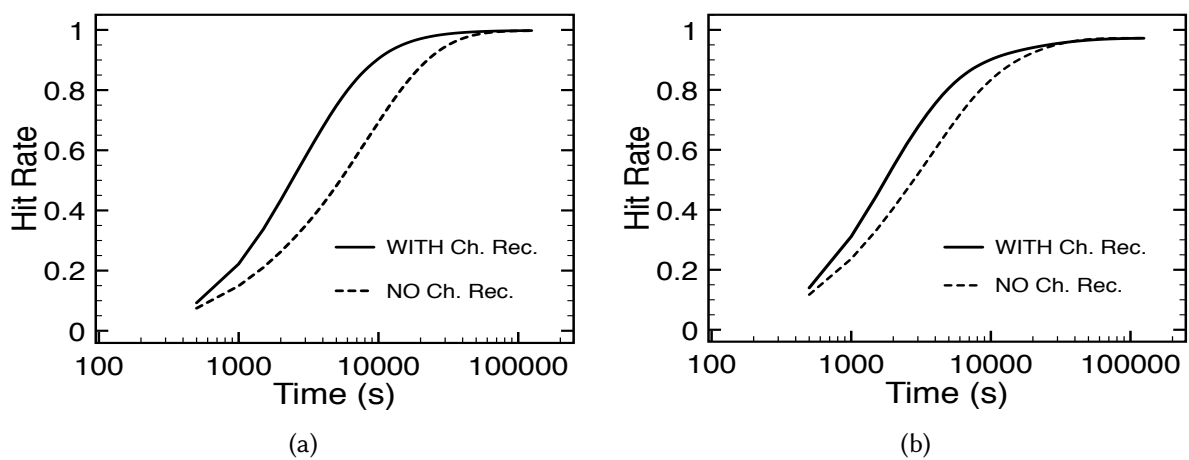

Fig. 12. Channel recognition impact: (a) Scenario 1; (b) Scenario 2. Only channels with subscribers on at least 25% of the communities

Performance results characterise the behaviour of the cognitive-based data dissemination along several directions, including number and popularities of data, number of users, geographical scales, and mobility patterns. The single most typical feature observed across the simulations is that the considered scheme is able to effectively disseminate data items across a wide range of popularities, while using a very limited amount of storage space at each node. This is very important, as serving efficiently content "in the long tail" is key, and not consuming individual nodes' resources is also fundamental. At a finer level of detail, our hybrid simulation model allows us to highlight differences due to the scheme and environment parameters. Quite interestingly, for example, we have found a phase transition, at the point where less than one node per social community is interested in data types (channels). For popularities below that point, as it is intuitive, the speed of dissemination drops dramatically, while differences are not so pronounced as soon as popularities increase over that threshold.

We can thus conclude that the analysed data dissemination scheme is very efficient in bringing data to interested users, and is able to support completely decentralised and self-organising data sharing schemes even up to the geographical scale of a region (or a small country), populated by a few million users.

## ACKNOWLEDGMENTS

This work is partly funded by the EC under the H2020 REPLICATE (691735) and AUTOWARE (723909) projects.

# REFERENCES

[1] [n.d.]. 5G Road Map Whitepaper. https://5g-ppp.eu/wp-content/uploads/2017/03/5GPPP-brochure-final-web-MWC.pdf.

[2] [n.d.]. Cisco VNI Mobile Forecast White Paper (2016Ð-2021), Feb. 2017. http://www.cisco.com/c/en/us/solutions/collateral/service-provider/visual-networking-index-vni/complete-white-paper-c11-481360.html.

[3] Anastasia Anagnostou, Athar Nouman, and Simon JE Taylor. 2013. Distributed Hybrid Agent-based Discrete Event Emergency Medical Services Simulation. In *2013 Winter Simulation Conference (WSC)*. IEEE, 1625–1636.

[4] P. Bellavista, S. Chessa, L. Foschini, L. Gioia, and M. Girolami. 2018. Human-Enabled Edge Computing: Exploiting the Crowd as a Dynamic Extension of Mobile Edge Computing. *IEEE Communications Magazine* 56, 1 (Jan 2018), 145–155. https://doi.org/10.1109/MCOM.2017.1700385

[5] Chiara Boldrini, Marco Conti, and Andrea Passarella. [n.d.]. Design and performance evaluation of ContentPlace, a social-aware data dissemination system for opportunistic networks. *Comput. Netw.* 54 ([n. d.]), 589–604. Issue 4.

[6] Chiara Boldrini and Andrea Passarella. [n.d.]. HCMM: Modelling spatial and temporal properties of human mobility driven by users' social relationships. *Comput. Commun.* 33 ([n. d.]), 1056–1074. Issue 9.

[7] Eleonora Borgia. 2014. The Internet of Things vision: Key features, applications and open issues. *Comp. Comm.* 54 (2014), 1–31.

[8] Raffaele Bruno, Marco Conti, Matteo Mordacchini, and Andrea Passarella. 2012. An analytical model for content dissemination in opportunistic networks using cognitive heuristics. In *Proc. of the 15th ACM international conference on Modeling, analysis and simulation of wireless and mobile systems (MSWiM 2012)*. ACM, 61–68.

[9] M. Conti and R. Mirandola. 1998. Hierarchical Performance Modeling of Computer Communication Systems. In *Network Performance Modeling and Simulation*, J. Walrand, K. Bagchi, and G. G.Zobrist (Eds.). Gordon and Breach Science Publishers, Chapter 8, 197–218.

[10] Marco Conti, Matteo Mordacchini, and Andrea Passarella. 2013. Design and performance evaluation of data dissemination systems for opportunistic networks based on cognitive heuristics. *ACM Trans. Auton. Adapt. Syst.* 8, 3 (2013), 12:1–12:32.

[11] Marco Conti, Matteo Mordacchini, Andrea Passarella, and Liudmila Rozanova. 2013. A Semantic-based Algorithm for Data Dissemination in Opportunistic Networks. In *Proc. of the 7th Int. Workshop on Self-organizing Systems (IWSOS 2013)*. Springer, 14–26.

[12] Marco Conti and Andrea Passarella. 2018. The Internet of People: A human and data-centric paradigm for the Next Generation Internet. *Computer Communications* 131 (2018), 51 – 65. COMCOM 40 years.

[13] Marco Conti, Andrea Passarella, and Sajal K Das. 2017. The Internet of People (IoP): A new wave in pervasive mobile computing. *Perv. and Mob. Comp.* 41 (2017), 1–27.

[14] Paolo Costa, Cecilia Mascolo, Mirco Musolesi, and Gian Pietro Picco. 2008. Socially-aware routing for publish-subscribe in delay-tolerant mobile ad hoc networks. *IEEE Journal on Selected Areas in Communications* 26, 5 (2008), 748–760.

[15] Gabriele D'Angelo, Stefano Ferretti, and Vittorio Ghini. 2018. Distributed Hybrid Simulation of the Internet of Things and Smart Territories. *Concurrency and Computation: Practice and Experience* 30, 9 (2018), e4370.

[16] H Ebbinghaus. 1913. Memory: A Contribution to Experimental Psychology Über das Gedchtnis.

[17] T. Eldabi, M. Balaban, S. Brailsford, N. Mustafee, R.E. Nance, B.S. Onggo, and R.G. Sargent. 2016. Hybrid Simulation: Historical lessons, present challenges and futures. In *2016 Winter Simulation Conference (WSC)*. IEEE, 1388–1403.

[18] G. Gigerenzer. 2004. Fast and frugal heuristics: The tools of bounded rationality. *Blackwell handbook of judgment and decision making* (2004), 62–88.

[19] G. Gigerenzer and D.G. Goldstein. 2011. The recognition heuristic: A decade of research. *Judgment and Decision Making* 6, 1 (2011), 100–121.

[20] Gerd Gigerenzer and Daniel G. Goldstein. 2002. Models of Ecological Rationality: The Recognition Heuristic. *Psych. Rev.* 109, 1 (2002), 75–90.

[21] Gerd Gigerenzer and Peter M Todd. 1999. Fast and frugal heuristics: The adaptive toolbox. In *Simple heuristics that make us smart*. Oxford University Press, 3–34.

[22] Daniel G. Goldstein and Gerd Gigerenzer. 1996. Reasoning the Fast and Frugal Way: Models of Bounded Rationality. *Psych. Rev.* 103, 4 (1996), 650–669.

[23] Daniel G Goldstein and Gerd Gigerenzer. 1999. The recognition heuristic: How ignorance makes us smart. In *Simple Heuristics That Make Us Smart*. Oxford University Press, 37Ð–58.

[24] Pan Hui, Jon Crowcroft, and Eiko Yoneki. 2011. Bubble rap: Social-based forwarding in delay-tolerant networks. *IEEE Transactions on Mobile Computing* 10, 11 (2011), 1576–1589.

[25] Thomas R. Kiehl, Robert M. Mattheyses, and Melvin K. Simmons. 2004. Hybrid Simulation of Cellular Behaviour. *Bioinformatics* 20, 3 (2004), 316–322.

[26] Vincent Lenders, Martin May, Gunnar Karlsson, and Clemens Wacha. [n.d.]. Wireless ad hoc podcasting. *ACM Mob. Comp. Comm. Rev.* 12 ([n. d.]), 65–67. Issue 1.

[27] K.C.-J. Lin, Chun-Wei Chen, and Cheng-Fu Chou. 2012. Preference-aware content dissemination in opportunistic mobile social networks. In *Proc. of IEEE INFOCOM 2012*. 1960–1968.
[28] Lauri LŁttilŁ, Per Hilletofth, and Bishan Lin. 2010. Hybrid simulation models Ð When, Why, How? *Expert Systems with Applications* 37, 12 (2010), 7969 – 7975.
[29] ÃĄlvaro Monares, Sergio F. Ochoa, Rodrigo Santos, Javier Orozco, and Roc Meseguer. 2014. Modeling IoT-Based Solutions Using Human-Centric Wireless Sensor Networks. *Sensors* 14, 9 (2014), 15687–15713.
[30] Matteo Mordacchini, Andrea Passarella, and Marco Conti. 2017. A social cognitive heuristic for adaptive data dissemination in mobile Opportunistic Networks. *Perv. and Mob. Comp.* (2017).
[31] Matteo Mordacchini, Andrea Passarella, Marco Conti, Stuart M. Allen, Martin J. Chorley, Gualtiero B. Colombo, Vlad Tanasescu, and Roger M. Whitaker. 2015. Crowdsourcing Through Cognitive Opportunistic Networks. *ACM Trans. Auton. Adapt. Syst.* 10, 2 (2015), 13:1,13:30.
[32] Matteo Mordacchini, Lorenzo Valerio, Marco Conti, and Andrea Passarella. 2016. Design and evaluation of a cognitive approach for disseminating semantic knowledge and content in opportunistic networks. *Comp. Comm.* 81 (2016), 12–30.
[33] Sergio F. Ochoa and Rodrigo Santos. 2015. Human-centric wireless sensor networks to improve information availability during urban search and rescue activities. *Information Fusion* 22 (2015), 71 – 84.
[34] J. Powell and N. Mustafee. 2014. Soft OR Approaches in Problem Formulation Stage of a Hybrid M&S Study. In *Proc. Of the Winter Simulation Conference*. IEEE, 1664–1675.
[35] Joshua Reich and Augustin Chaintreau. 2009. The age of impatience: optimal replication schemes for opportunistic networks. In *Proc. of ACM CoNEXT '09*. ACM, New York, NY, USA, 85–96.
[36] P. K. Reichl. 2013. It's the Ecosystem, Stupid: Lessons from an Anti-Copernican Revolution of User-Centric Service Quality in Telecommunications. In *2013 Sixth International Conference on Developments in eSystems Engineering*. 163–168.
[37] D. Rosário, M. Seruffo, E. Cerqueira, C. Both, T. Braun, and M. Gerla. 2016. Trends in Human-Centric Multimedia Networking scenarios. In *2016 Mediterranean Ad Hoc Networking Workshop (Med-Hoc-Net)*. 1–5. https://doi.org/10.1109/MedHocNet.2016.7528428
[38] L.J. Schooler and R. Hertwig. 2005. How forgetting aids heuristic inference. *Psych. Rev.* 112, 3 (2005), 610.
[39] H. D. Schwetman. 1978. Hybrid Simulation Models of Computer Systems. *Commun. ACM* 21, 9 (1978), 718–723.
[40] J. G. Shanthikumar and R. G. Sargent. 1983. A Unifying View of Hybrid Simulation/Analytic Models and Modeling. *Operations Research* 31, 6 (1983), 1030–1052.
[41] Mani Srivastava, Tarek Abdelzaher, and Boleslaw Szymanski. 2012. Human-centric sensing. *Philosophical Transactions of the Royal Society A: Mathematical, Physical and Engineering Sciences* 370, 1958 (2012), 176–197.
[42] Lorenzo Valerio, Andrea Passarella, Marco Conti, and Elena Pagani. 2015. Scalable data dissemination in opportunistic networks through cognitive methods. *Perv. and Mob. Comp.* 16 (2015), 115–135.
[43] Alberto Sangiovanni Vincentelli. 2015. Let's get physical: Adding physical dimensions to cyber systems. In *IEEE/ACM Int. Symp. on Low Power Electronics and Design (ISLPED)*. IEEE, 1–2.
[44] Jiangtao Wang, Junfeng Zhao, Yong Zhang, Xin Peng, Ying Li, and Yun Xie. 2018. Enabling Human-Centric Smart Cities: Crowdsourcing-Based Practice in China. *Computer* 51, 12 (2018), 42–49.
[45] John T Wixted and Ebbe B Ebbesen. 1991. On the form of forgetting. *Psychological science* 2, 6 (1991), 409–415.
[46] Eiko Yoneki, Pan Hui, ShuYan Chan, and Jon Crowcroft. 2007. A socio-aware overlay for publish/subscribe communication in delay tolerant networks. In *MSWiM*. 225–234.